\documentclass[11pt,twoside]{article}

\usepackage{baaa}

\usepackage{graphicx}
\usepackage{subfigure}
\usepackage{amssymb}
\usepackage[spanish,activeacute,english]{babel}
\usepackage[latin1]{inputenc}
\usepackage{verbatim}
\usepackage{amsmath}
\usepackage{amsfonts}
\usepackage{amssymb}
\usepackage[colorlinks=true,dvips]{hyperref}
\usepackage{natbib}

\begin{document}
%%%%%%%%%%%%%%%%%%%%%%%%%%%%%%%%%%%%%%%%%%%%%%%%%%%%%%%%%%%%%%%%%%%%%%%%%%
%%%% SELECCIONE EL IDIOMA EN QUE SE ESCRIBE EL ARTÍCULO:              %%%%
%\myselectspanish
\myselectenglish
%%%%%%%%%%%%%%%%%%%%%%%%%%%%%%%%%%%%%%%%%%%%%%%%%%%%%%%%%%%%%%%%%%%%%%%%%%
\vskip 1.0cm
\markboth{De Rossi et al.}%
{}

\pagestyle{myheadings}
%%%% DESCOMENTE LA LINEA QUE DESCRIBE EL CARACTER DE SU TRABAJO       %%%%
\vspace*{0.5cm}

%\noindent TRABAJO INVITADO 
%\noindent PRESENTACIÓN ORAL
\noindent PRESENTACIÓN MURAL
%\noindent RESUMEN 

\vskip 0.3cm
\title{Downsizing of galaxies vs upsizing of dark-halos in a $\Lambda$-CDM cosmology}

%\title{ Template paper for publication in the Bulletin of the 
%Argentinian Astronomical Association with instructions for the use of 
%\LaTeX{}}

\author{M.E. De Rossi$^{1,2,3}$, V. \'Avila{\--}Reese$^{4}$, A. Gonz\'alez-Samaniego$^{4}$ \& S.E. Pedrosa$^{1,2}$}

\affil{%
  (1) Instituto de Astronomía y Física del Espacio (CONICET-UBA)\\ 
  (2) Consejo Nacional de Investigaciones Cient\'ificas y T\'ecnicas, CONICET, Argentina (derossi@iafe.uba.ar)\\
  (3) Facultad de Ciencias Exactas y Naturales, Universidad de Buenos Aires, Ciudad Aut\'onoma de Buenos Aires, Argentina\\
  (4) Instituto de Astronom\'ia, Universidad Nacional Aut\'onoma de M\'exico, A.P- 70-264, 04350 M\'exico, D.F., M\'exico\\
}

\begin{abstract} 
The mass assembly of a whole population of sub-Milky Way galaxies
is studied by means of hydrodynamical simulations
within the $\Lambda$-CDM cosmology.  Our results show that
while dark halos assemble hierarchically, in stellar mass this trend
is inverted in the sense that the smaller the galaxy, the later is its
stellar mass assembly on average. Our star formation and supernovae feedback
implementation in a multi-phase interstellar medium seems to play a key role on
this process. However, the obtained downsizing trend is not yet
as strong as observations show.
\end{abstract}

\begin{resumen}
El ensamblaje de la masa de una población completa de galaxias {\it sub-Milky Way}
es estudiada por medio de simulaciones hidrodinámicas dentro de una cosmología $\Lambda$-CDM.
Nuestros resultados muestran que mientras los halos oscuros se ensamblan jerárquicamente,
esta tendencia se invierte en la masa estelar en el sentido de que cuánto más pequeña es la
galaxia, su ensamblaje es más tardío en promedio.  Nuestra implementación de la
formación estelar y retroalimentación al medio por supernovas en un medio
interestelar multi-fase parace jugar un rol clave en este proceso.  Sin embargo, la tendencia
de {\it downsizing} obtenida no es aún tan fuerte como muestran las observaciones.
\end{resumen}

\section{Introduction}
\label{intro}

In the hierarchical $\Lambda$-Cold Dark Matter ($\Lambda$-CDM) scenario, 
the cold dark matter halos assemble on average first 
the smaller ones, and successively the larger ones (upsizing). 
In this context, it is interesting to study whether the assembly of baryons and stars in galaxies
exhibit similar trends to their halos.
In fact, galaxy formation process implies much more physical mechanisms than the gravitational one 
(e.g., gas heating, dissipation, and infall, star formation and its feedback, 
active galaxy nuclei -AGN- feedback, among others), which on its own depend on halo mass and environment. 

From the observational point of view, there are increasing pieces of evidence that the specific 
star formation rates (sSFRs) of low-mass galaxies are very high since $z \sim 2$ in the sense 
that they imply star formation rates (SFRs) much higher than their corresponding past average SFR; 
besides the less massive the galaxy, the higher on average its sSFRs (downsizing in sSFR; 
for a recent review and more references see Avila-Reese \& Firmani 2011). 
These results seem to imply that the smaller the galaxy, the later it had on average its active 
onset of SF and the faster is its late stellar mass ($M_{*}$) assembly history (MAH).

By means of full N-body $+$ hydrodynamic cosmological simulations, we attempt to explore here how are 
the baryonic and stellar mass assembly of sub-MW galaxies  (Milky Way-sized and smaller galaxies) 
and how do compare these assembly histories with those of their halos as well as with observational inferences.

\section {Numerical simulations and galaxy sample}
We performed numerical simulations consistent with the concordance $\Lambda$-CDM universe
with $\Omega_m =0.3, \Omega_\Lambda =0.7, \Omega_{b} =0.045$, a normalisation of the power
spectrum of ${\sigma}_{8} = 0.9$ and $H_{0} =100 \, h$ km s$^{-1} \ {\rm Mpc}^{-1}$ with  $h=0.7$. 
These {simu}-{lations} were performed by using the chemical code GADGET-3 (Scannapieco et al. 2008), 
which includes treatments for metal-dependent radiative cooling, stochastic star formation, 
chemical enrichment and  supernovae feedback.  No AGN feedback is included in our simulation. 

The simulated volume corresponds to a cubic box of a comoving 10 Mpc $h^{-1}$ side length. 
We analysed two simulations. The masses of dark matter and initial gas-phase particles are 
$5.93 \times 10^6 M_{\odot} h^{-1}$ and  $9.12 \times 10^5 M_{\odot} h^{-1}$, respectively (S230) 
and $2.20 \times 10^6 M_{\odot} h^{-1}$ and  $3.40 \times 10^5 M_{\odot} h^{-1}$, respectively (S320).  
Due to computational costs, we were only able to run S320 up to z=2.  
See De Rossi et al. (2010, 2012a) for more details about these simulations.

\section{Results and discussion}

\begin{figure}[!ht]
  \centering
  \includegraphics[width=1\textwidth]{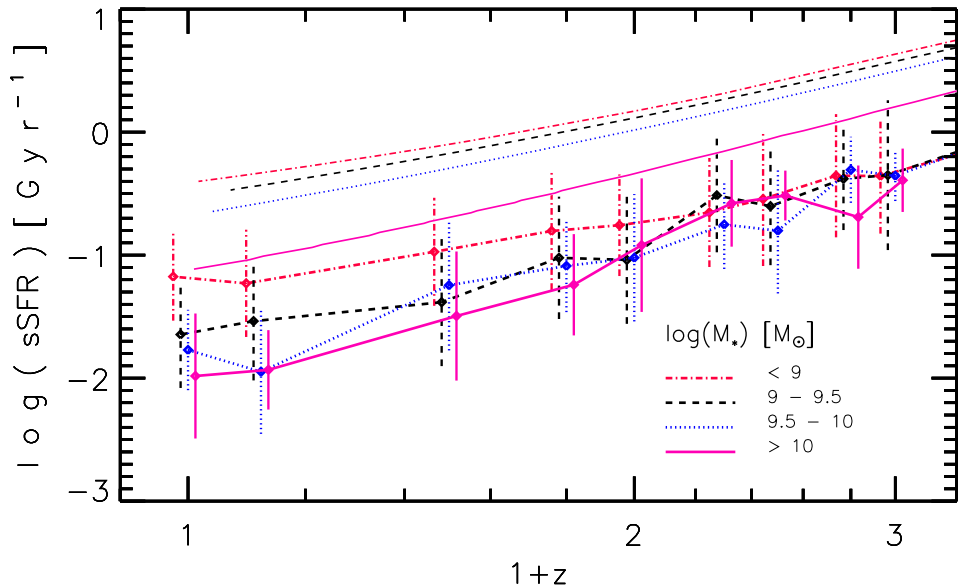}
  \caption{
Mean sSFR versus $1+z$ for different fixed bins of $\log (M_{*} / M_{\odot})$:
$< 9$ (dotted-dashed red line), $9 - 9.5$ (dashed black line), $9.5 - 10$ (dotted blue line), $> 10$ (solid pink line).
The vertical lines denote the standard deviations associated to simulated values.  The curves without
standard deviations correspond to results from the parametric model tuned to fit 
observational inferences (see De Rossi et al. 2012b).
}
  \label{fig:sSFR_vs_z}
\end{figure}

In Fig. \ref{fig:sSFR_vs_z} we show the change with redshift
of the average simulated sSFR at different masses: for fixed bins of $M_{*}$ -the same ones at all epochs-,
the average sSFR and its standard deviation are calculated at different epochs and plotted versus
$z$ (note that these results do not refer to individual evolutionary tracks -discussed in De Rossi et al. 2012b- 
but to population averages at different epochs). From Fig. \ref{fig:sSFR_vs_z}, we can appreciate that:
(a) the average overall sSFR of the simulated galaxy population decreases significantly as $z$ decreases,
a factor of $\sim 15$ since $z=2$ to $z=0$;
(b) since $z\sim 2$, at a given $z$, the less massive galaxies have higher sSFRs on average than
more massive galaxies being this difference higher for lower $z$.

The curves without standard deviations plotted in Fig. \ref{fig:sSFR_vs_z}
correspond to results from a parametric model of SFR and $M_{*}$ evolution specially tuned to fit
observational inferences (see De Rossi et al. 2012b). 
Note that simulated galaxies have
lower sSFRs on average than the trends from observational inferences, this difference
increasing as $z$ is lower. This means that the assembly of $M_{*}$ by in situ SF of simulated
galaxies tends to happen earlier on average than what observations would suggest. On the other hand,
our simulation results follow qualitatively the observed trend of higher sSFR as
smaller is $M_{*}$ at low redshifts (downsizing in sSFR) and the flattening of this correlation
for higher redshifts.

As analysed in detail by De Rossi et al. (2012b), for both the Millennium simulations and our one,
massive systems assemble their virial masses on average slightly later than the less massive ones 
(upsizing), though in our case this trend is less pronounced. 
By comparing the average halo-to-galaxy stellar MAHs, we found that the upsizing
trend of the halos tends to be reversed on average to a downsizing trend for $M_{*}$ in the sense that 
galaxies in the less massive halos tend to assemble their present-day $M_{*}$ later than those in 
the massive ones, on average.  
In particular, 
the redshift at which simulated $M_{*}$ attains 50\% of its present-day value is on average
$z_{\rm h,1/2}=0.8$ for halo masses $\log(M_{\rm vir}/M_{\odot}) <10.5$,
and $z_{\rm h,1/2}= 1.3$ for halo masses $\log(M_{\rm vir}/M_{\odot}) > 11.5$.
The reader is referred to De Rossi et al. (2012b) for more details and figures.

With respect to the baryonic mass assembly history as a function of mass, the trends 
do not differ significantly on average from the stellar one, i.e. 
the downsizing trend with stellar mass is roughly the same for the baryonic mass. 
Finally, the average stellar mass growth tracks inferred with a toy model of stellar mass assembly 
inside growing cold dark matter halos and constrained to reproduce the empirical sSFR($M_{*}$, $z$) 
and $M_{*}$ ($M_{\rm vir}$, $z$) relations, show a stronger downsizing than simulations. 
These observations-based $M_{*}$ tracks evolve lately much faster than those measured in simulations.

\section{Conclusions}

The stellar and baryonic mass assembly of sub-MW galaxies 
are studied simultaneously with the hierarchical mass aggregation of their parent dark-matter 
halos by using cosmological simulations. 
Consistently with previous works, the dark halos assemble hierarchically in the sense that 
low-mass systems acquire most of their mass earlier than more massive ones.

The astrophysical processes included in  these simulations 
seem to be able to partially invert this behavior in 
that regards stellar mass assembly.  However,  this trend is still weak as compared to empirical 
inferences. Our sub-MW simulated galaxies (1) have specific SFRs at 
late epochs on average lower than those observed, and (2) their stellar mass fractions 
are higher than semi-empirical inferences, specially at larger $z$.
As discussed in detail in De Rossi et al. (2012b),  the efficiency of the current
SN feedback model could hardly be increased. Probably, the inclusion of
treatments for other feedback mechanisms
(stellar winds, HII photoionization, etc.) 
can contribute to tackle the problem. It is also important to revise the modelization
of the star formation process.  In particular, the formation of molecular gas $H_2$ 
could be relevant.

\section*{Acknowledgements}

We thank the anonymous referee for his/her useful comments that
helped to improve this paper.
We acknowledge a CONACyT-CONICET (M\'exico-Argentina) bilateral grant for partial funding.
V.A. and A. G. acknowledge PAPIIT-UNAM grant IN114509.  A.G. acknowledges a PhD fellowship
provided by CONACyT.
M.E.D.R. and S.P. acknowledge support from the  PICT 32342 (2005),
PICT 245-Max Planck (2006) of ANCyT (Argentina), PIP 2009-112-200901-00305 of
CONICET (Argentina) and the L'oreal-Unesco-Conicet 2010 Prize.
Simulations were run in Fenix and HOPE clusters at IAFE and Cecar cluster at University
of Buenos Aires, Argentina.

\begin{referencias}

\reference Avila-Reese, V., \& Firmani, C.\ 2011, Revista Mexicana de Astronomia y Astrofisica Conference Series, 40, 27 

\reference de Rossi, M.~E., Tissera, P.~B., \& Pedrosa, S.~E.\ 2010, \aap, 519, A89

\reference De Rossi, M.~E., Tissera, P.~B., \& Pedrosa, S.~E.\ 2012a, \aap, 546, A52 

\reference De Rossi, M.~E., Avila-Reese, V., Tissera, P.~B., A., Gonz\'alez-Samaniego \& Pedrosa, S.~E., 2012b, submitted

\reference Scannapieco, C., Tissera, P.~B., White, S.~D.~M., \& Springel, V.\ 2008, \mnras, 389, 1137

\end{referencias}

\end{document}